\documentclass[
reprint,
aps,
prb,
]{revtex4-2}

\usepackage{graphicx}
\usepackage{dcolumn}
\usepackage{bm}
\usepackage{bbm}
\usepackage{amsmath, amssymb}
\usepackage[dvipsnames]{xcolor}

\begin{document}

\preprint{APS/123-QED}

\title{General and concise operator approach to the dyadic Green's function of layered media}

\author{Aliaksandr Arlouski$^{1}$}
 \email{av.arlouski@gmail.com}
 \author{Lei Gao$^{2,3}$}
\author{Dongliang Gao$^{4}$}
\author{Andrey Novitsky$^{1}$}
 \email{andreyvnovitsky@gmail.com}

\affiliation{$^1$ Belarusian State University, Nezavisimosti Avenue 4, 220030 Minsk, Belarus
\\ $^2$ School of Optical and Electronic Information, Suzhou City University, Suzhou 215104, China
\\ $^3$ Jinagsu Key laboratory of Biophotonics \& Suzhou Key Laboratory of Biophotonics \& SZU-ITU International Joint Meta-center for Advanced Photonics and Electronics, Suzhou City University, Suzhou 215104, China
\\ $^4$ School of Physical Science and Technology \& Jiangsu Key Laboratory of Frontier Material Physics and Devices, Soochow University, Suzhou 215006, China}

\date{\today}

\begin{abstract}
Dyadic Green's function is an important tool of computational photonics, giving deeper insights into light-matter interaction. We present an operator approach to the derivation of the dyadic Green's function of a generic anisotropic planarly-layered medium for both electric and magnetic fields. The resulting Green's function is expressed through the evolution operators (a kind of transfer matrices) of the comprising layers and the surface impedance tensors, the singular term being naturally separated from other terms. The operator approach to the Green's function simplifies both the conceptual understanding of the problem and the subsequent practical applications, some of which are demonstrated here. The proposed approach can be easily generalized to the case of spherical and cylindrical layers, as well as bi-anisotropic layered media. The obtained results can be applied in nanophotonics engineering problems.
\end{abstract}

\maketitle


\section{\label{sec:level1}Introduction}

The Green’s function is a solution to a partial differential equation with a point source, allowing one to find a solution of the corresponding inhomogeneous equation satisfying boundary conditions \cite{kuzemsky2017,rother2017}. When dealing with classical electrodynamics, the equations at hand are Maxwell's equations, the unknown quantities being the electric and magnetic (EM) fields. The Green's function for obtaining the distribution of these fields in space using the known distribution of electric currents is called the dyadic Green's function (DGF), which is, in general, not a scalar, but a tensor function \cite{tai1994,faryad2018}.

DGF is ubiquitous in photonics, since it contains all the information about the system under study necessary to obtain the distribution of the electromagnetic field in space: It encapsulates the geometry of the environment, along with all of its relevant material properties. That is, finding the DGF is tantamount to solving the Maxwell's equations for a specific environment. Having obtained it, one can immediately calculate the distribution of the electromagnetic field due to arbitrary electric currents, as well as various useful quantities, such as the local density of optical states (LDOS) \cite{barnes2020}. Diverse applications of the DGF have been proposed, which include the determination of stationary phase points \cite{luomei2024}, Purcell factor enhancement \cite{bordo2012,jain2024, ren2024}, and optimization of layered media to get the desired response \cite{qi2023}, to name a few.

It is easy to derive the DGF for homogeneous media \cite{novotny2012, chew1999}. However, this case is of little interest for practical purposes: any nontrivial environment scatters and/or absorbs electromagnetic radiation. In that case, the derivation of the DGF may become exhausting: the amount of bookkeeping necessary quickly increases with the addition of new components to the environment. This issue is clearly manifest when finding the DGF of planarly-layered media, which are basic environments both in theory and practice. Even in the case of just two adjacent semi-infinite isotropic layers, the calculations required are quite extensive \cite{cho2016}.

Many techniques and approximations for finding the DGF of various isotropic layered media have been developed (see \cite{paulus2000, cho2016, sipe1987, chew1999, hohenester2021, koh2024, lambot2024,ali1992,okhmatovski2024} and references therein). A pedagogical discussion of general properties of DGFs can be found in \cite{novotny2012, hohenester2021}. A detailed treatment of a bi-layered medium is given in \cite{cho2016, koh2022}, where the authors also discuss how to implement numerical calculations based on their work. In \cite{chew1999, sipe1987, paulus2000} one can find general algorithms for finding the DGF of a layered medium with an arbitrary number of isotropic layers, and in \cite{ali1992} authors deal with the DGF of a layered medium with chiral layers. 

There are also many works that consider anisotropic media, both layered and unbounded \cite{Barkeshli1993, Krowne1986, Tsalamengas1992, LosethUrsin2007, TanTan1998, TanTan2000, TanTan2001, Li2004, HuangLee2011, Wei2008,potemkin2012}. In \cite{potemkin2012}, an unbounded hyperbolic medium is treated; however, it is not discussed how to apply the interface conditions to the obtained result in order to deal with the multilayer case. In \cite{Barkeshli1993}, the symmetric gyroelectric anisotropy of permittivity $\varepsilon$ of layers is considered. The treatment of an unbounded or two-layered medium with general anisotropic $\varepsilon$ tensor and can be found in \cite{HuangLee2011}. In \cite{Li2004}, the permittivity is taken to be a gyroelectric uniaxial tensor. In \cite{Wei2008}, authors consider stratified uniaxial anisotropic media. Note that in all the works above, the magnetic permeability $\mu$ of layers is taken to be a scalar. In \cite{LosethUrsin2007}, authors assume that $\varepsilon$ and $\mu$ are symmetric tensors, which is achieved by rotating the coordinate axes to principal axes of material parameters tensors. There are also papers that do not make any assumptions about the form of the $\varepsilon$ and $\mu$ tensors of layers: in \cite{Tsalamengas1992}, \cite{Krowne1986} and \cite{TanTan2001} (see also their prior works \cite{TanTan1998, TanTan2000}) authors treat the most general bi-anisotropic layered media.

The methods to obtain the DGF of anisotropic and bi-anisotropic media vary. In \cite{Barkeshli1993}, the $z$-propagating eigenmode expansion and multiple scattering method (MSM) are employed. \cite{HuangLee2011} uses the eigen-decomposition method. \cite{Li2004} uses eigenfunction expansion method with cylindrical vector wavefunctions based on the principle of scattering superposition. In \cite{Wei2008}, Sommerfeld integral representation of vector potential is used, the $z$-components of electric and magnetic fields are expressed as superpositions of upward and downward propagating waves within each layer, and a recursive algorithm is used to solve matrix equations. Another use of a recursive algorithm can be found in \cite{LosethUrsin2007}, where it is used to find reflection and transmission coefficients for a stack of layers. In \cite{TanTan2001}, authors decompose the DGF into unbounded and scattered parts, the first of which is decomposed in terms of vector wave functions, and the second is constructed using scattering coefficient matrix; same authors previously used a distribution-theoretic approach combined with eigenfunction expansion \cite{TanTan1998}. And in \cite{Krowne1986, Tsalamengas1992} authors employ variations of the $4 \times 4$ matrix method in spectral domain and transfer matrix approach (see also the foundational paper \cite{Berreman1972}, where the $4 \times 4$ matrix method for anisotropic media is introduced).

In the present work, we derive the DGF of a generic anisotropic planarly-layered medium in an operator-based way \cite{fedorov1958,fedorov1976,Borzdov97,barkovsky_furs2003}. That is, we try to make our derivations coordinate-free, treating vectors and operators (tensors) as geometric objects, without regard to their components in some selected basis. This method has several advantages compared to the approaches mentioned above.

First, the derivation of the DGF using this approach is clearer and more concise, since it makes it unnecessary to treat the DGF as a matrix in some fixed basis and then find its expression component-by-component, thus avoiding lengthy algebraic calculations. Also, we do not have to clutter out notation with redundant indices for vector and tensor components.

Next, the singular term of the DGF (the one containing the delta-function) should be properly treated and separated from other terms, since calculation of, e.g., LDOS requires the evaluation of the Green's function at $\mathbf{r} = \mathbf{r}'$. Among papers that explicitly mention this point are \cite{Barkeshli1993, Li2004, TanTan2001, LosethUrsin2007, Tsalamengas1992}, with \cite{TanTan1998, TanTan2000, Barkeshli1993} being especially thorough about it. In our work, the singular term is naturally separated after performing an inverse Fourier transform. This is a direct consequence of the operator method, which allows to avoid extra steps to separate the singularity, required by some earlier approaches \cite{Barkeshli1993, Li2004, potemkin2012}. We also find that the source term is only present in the electric part of the DGF, which is expected, since there are no magnetic point charges, and that the coefficient in front of the source term does not depend on the magnetic permeability of layers.

When deriving the DGF (which is usually taken to be the DGF for the electric field), it would be beneficial to simultaneously derive the magnetic Green's function, so that both electric and magnetic fields could be treated in a similar manner. Of course, the two DGFs can be related to each other via the Maxwell's equations, and if one is known, the other can also be obtained. But since we want the resulting DGF to be ready to use for practical application, it would be expedient to also present an expression for the magnetic DGF. Some works do treat both electric and magnetic DGFs \cite{Barkeshli1993, Wei2008, TanTan2001, LosethUrsin2007, Tsalamengas1992}, while others do not \cite{HuangLee2011, Li2004, Krowne1986}. In our approach, both of the Green's functions are treated in exactly the same way, and no additional steps are required to obtain the magnetic DGF.

There are few papers where authors discuss how or if their methods can be generalized for non-planar geometries or non-Cartesian coordinate systems in the case of anisotropic or bi-anisotropic media. One such paper is \cite{TanTan1998}; however, authors deal with an unbounded bi-anisotropic medium, not a multi-layered one. On the other hand, one of the main benefits of the operator method lies in how it can be readily used to treat non-planar layers. As we discuss in Appendix \ref{app:4}, the form of the main equation of our paper and its solution (Eqs. (\ref{basicEq}) and (\ref{homogeneousSolution}) respectively) keep their form for spherical and cylindrical geometries, and we give the form of the source term for such shape of layers. The only novelty would be that the $M$ operator coefficient in Eq. (\ref{basicEq}) would now depend on the stratification parameter, which would require some extra effort when numerically evaluating the evolution operator $\Omega$. But the general algorithm still remain the same for non-planar geometries. It is worth to mention that our approach allows for an easy treatment of bi-anisotropic media as well: in that case, Eqs. (\ref{basicEq}) and (\ref{homogeneousSolution}) also remain their form. So, the operator approach does not restrict the possible media to anisotropic media only and allows one to treat the DGF for more complicated bi-anisotropic media in the same manner.

Finally, while the DGFs for isotropic media have been discussed in great details in textbooks (see, e.g., \cite{novotny2012} and especially \cite{chew1999}), the pedagogical treatment of anisotropic and bi-anisotropic layered media is much harder to come by (a hint on how to apply the formalism of the DGF to anisotropic media is given in Section 8.1.3 of \cite{chew1999}; however, no final expression for such a DGF is given). Among the papers that we have mentioned, we can especially point out \cite{TanTan2001, Tsalamengas1992, LosethUrsin2007}, where authors try to make every step of the derivation clear. So, a paper about a DGF for complex media should, ideally, be easy to follow even to those who do not have a big experience in this matter (e.g., experimentalists, who just wish to use the obtained DGF for their purposes). In our work, we have tried to make sure that readers unfamiliar with the operator method could follow every step of our derivation without much effort. We discuss various aspects regarding our derivation, like the similarity of Eq. (\ref{basicEq}) to the Schrödinger's equation of quantum mechanics, and also discuss how the result obtained using our approach can be implemented for practical numerical calculations by choosing an appropriate coordinate system in which, e.g., the permittivity tensors of the comprising layers have the simplest form. We perform several such calculations and give some tips on how to conduct them efficiently.

The structure of the paper is as follows. In Section \ref{sec:2}, we derive the DGF of a planarly-layered medium, trying to present the derivation in a manner which would allow the reader unfamiliar with the operator approach to easily comprehend it. The mathematical details can be found in Appendices \ref{app:1} and \ref{app:2}. 
After that, we discuss some neat points regarding our derivation. In Section \ref{sec:3}, numerical calculations with the obtained DGF are performed. In particular, we calculate the LDOS enhancement for several layered media and discuss how one can efficiently implement them. Finally, in Section \ref{sec:4} we briefly summarize the work done, and in Appendix \ref{app:4} we explain how the obtained results can be generalized to cylindrical and spherical layers.

\section{\label{sec:2}Deriving dyadic Green's function of a planarly-layered medium}

The most common approach to obtaining the DGF of multilayered media is as follows \cite{paulus2000, chew1999, novotny2012, hohenester2021, HuangLee2011, Li2004, TanTan2001}. First, from Maxwell's equations, one obtains a vector wave equation for the electric field $\mathbf{E}(\mathbf{r})$ of the form
\begin{equation} \label{vectorWaveEq}
    \nabla \times \nabla  \times \mathbf{E} - k^2\mathbf{E}  = i\mu \omega \mathbf{j}, ~\quad k^2 = \varepsilon \mu \dfrac{\omega^2}{c^2}. 
\end{equation}
This equation can be solved using its Green's function
\begin{equation} \label{usualHomDGF}
    G(\mathbf{r}, \mathbf{r}') = \left( 1 + \dfrac{\nabla \otimes \nabla}{k^2} \right) \dfrac{e^{ik|\mathbf{r} - \mathbf{r}'|}}{4 \pi |\mathbf{r} - \mathbf{r}'|}.
\end{equation}
The obtained solution is valid for a single homogeneous medium with isotropic (scalar) $\varepsilon$ and $\mu$. After that, the Fourier transform is applied to $G$, and then the DGF for the whole layered medium is found as the superposition of homogeneous Green's functions, each representing reflected and transmitted waves, with coefficients that are expressed through the reflection ($r$) and transmission ($t$) coefficients. 

Although perfectly valid and efficient for many special cases, this approach has several disadvantages. First, the Green's function (\ref{usualHomDGF}) only solves (\ref{vectorWaveEq}) if $\varepsilon$ and $\mu$ are scalars, and it is not obvious how to generalize (\ref{usualHomDGF}) to the anisotropic case. Second, if some of the layers are anisotropic, one cannot simply use the Fresnel's formulae for $r$ and $t$ coefficients, and finding them may become a quest on its own.

To mitigate such difficulties, our strategy for deriving the DGF is different. First, from the beginning we turn Maxwell's equations into equations for tangential components of the Fourier amplitudes of EM fields. It is convenient, since tangential components are continuous across interfaces of layers. After finding the EM field inside a single homogeneous medium with constant material parameters for a given distribution of electric currents $\mathbf{j}(\mathbf{r})$, we apply the interface conditions to find the EM field at any point both inside and outside the layered medium. Finally, we compare the resulting expression for the electric field $\mathbf{E}$ with the definition (up to a constant factor, as we later discuss) of the electric DGF
\begin{equation} \label{DGFdef}
    \mathbf{E}(\mathbf{r}) = \int \hat{G}_{E}(\mathbf{r}, \mathbf{r}') \mathbf{j}(\mathbf{r}') \mathrm{d}^3 \mathbf{r}',
\end{equation}
after which we are able to immediately deduce the form of $\hat{G}_{E}(\mathbf{r}, \mathbf{r}')$, along with the DGF for the magnetic field. In other words, we directly solve the Maxwell's equations for the layered medium under investigation, and then compare the results with (\ref{DGFdef}). Note that although the definition (\ref{DGFdef}) differs from the more common one, wherein $\hat{G}_{E}$ is defined via the electric current density produced by an oscillating dipole \cite{novotny2012}, these definitions are actually equivalent up to a constant factor, as we discuss later.

\subsection{Electromagnetic field in a homogeneous medium}

Assuming stationarity of the system, i.e. the time dependence of the fields and currents is only due to the factor of $e^{-i \omega t}$, the $\operatorname{curl}$ Maxwell's equations take the following form:
\begin{equation} \label{MaxwellEqs}
    \begin{cases}
		\nabla \times \mathbf{H}(\mathbf{r}) = -i k_0 \mathbf{D}(\mathbf{r}) + \dfrac{4\pi}{c} \mathbf{j}(\mathbf{r}), \\
		\nabla \times \mathbf{E}(\mathbf{r}) = i k_0 \mathbf{B}(\mathbf{r}),
	\end{cases}
\end{equation}
where $k_0 \equiv \omega/c$ is the wavenumber in vacuum and $\mathbf{j}$ is the electric current density generating the fields. According to the constitutive relations, the displacement field $\mathbf{D}$ and magnetic induction $\mathbf{B}$ are related to the field strengths as $\mathbf{D} = \varepsilon \mathbf{E}$ and $\mathbf{B} = \mu \mathbf{H}$.
We do not make any assumptions about the specific form of $\varepsilon$ and $\mu$: they are generic, non-Hermitian and non-symmetric tensors.

The reason why we do not need the $\operatorname{div}$ Maxwell's equations for all the derivations to follow lies in the fact that the $\operatorname{curl}$ equations, together with the continuity equation
\begin{equation}
    \dfrac{\partial \rho}{\partial t} + \nabla \cdot \mathbf{j} = 0,
\end{equation}
(which is just the law of conservation of electric charge and is, of course, assumed to hold) are tantamount to the full system of four Maxwell's equations. One can see this by taking the divergence of both equations in (\ref{MaxwellEqs}).

Assume that the layered medium is orthogonal to some unit vector $\mathbf{z}$, and $z \equiv \mathbf{r} \cdot \mathbf{z}$ is a parameter along the stratification axis aligned with $\mathbf{z}$. We perform the Fourier transform of the EM fields and of electric current density in a plane orthogonal to $\mathbf{z}$ (also known as the angular spectrum representation of fields \cite{novotny2012}):
\begin{equation}
		\mathbf{H}(\mathbf{r}) = \int \tilde{\mathbf{H}}(\mathbf{b}, z) e^{i k_0 \mathbf{b} \mathbf{r}_{t}} ~ \mathrm{d}^2(k_0^2 \mathbf{b}),
\end{equation}
same for $\mathbf{E}$ and $\mathbf{j}$, where $\mathbf{r}_{t}$ is the tangential 
\footnote{Here and below, “tangential” means “orthogonal to $\mathbf{z}$”.}
component of the radius vector, $\mathbf{b}$ is the dimensionless vector, such that $k_0 \mathbf{b}$ is the tangential component of the wavevector $\mathbf{k}_t$, and $\mathrm{d}^2 (k_0^2 \mathbf{b}) \equiv \mathrm{d}(k_0 b_1) \mathrm{d}(k_0 b_2) = \mathrm{d}^2 \mathbf{k}_t$. Physically, this means that we represent the fields as the sum of plane waves with different amplitudes and tangential wavevectors $\mathbf{k}_t$, that is, with different values of the angle of incidence. We denote Fourier amplitudes with tilde $\tilde{} ~$, and it is always implied that Fourier amplitudes depend on $\mathbf{b}$ and $z$, even if this dependency is not written explicitly. Since we assume that the properties of the medium are independent of the tangential coordinates, and since the Fourier transform is linear, the relations between $\tilde{\mathbf{D}}, \tilde{\mathbf{B}}$ and $\tilde{\mathbf{E}}, \tilde{\mathbf{H}}$ are the same as between $\mathbf{D}, \mathbf{B}$ and $\mathbf{E}, \mathbf{H}$, i.e. $\tilde{\mathbf{D}} = \varepsilon \tilde{\mathbf{E}}$ and $\tilde{\mathbf{B}} = \mu \tilde{\mathbf{H}}$. 

Upon substituting the Fourier-transformed fields into Eq. (\ref{MaxwellEqs}), we arrive at the following system of equations:
\begin{equation} \label{FourierAmpsEqs}
    \begin{cases}
		\left( \mathbf{z}^{\times} \dfrac{\mathrm{d}}{\mathrm{d} z} + i k_0 \mathbf{b}^{\times} \right) \tilde{\mathbf{H}} + i k_0 \varepsilon \tilde{\mathbf{E}} = \dfrac{4\pi}{c} \tilde{\mathbf{j}}, \\
        \left( \mathbf{z}^{\times} \dfrac{\mathrm{d}}{\mathrm{d} z} + i k_0 \mathbf{b}^{\times} \right) \tilde{\mathbf{E}} - i k_0 \mu\tilde{\mathbf{H}} = 0.
	\end{cases}
\end{equation}
Symbol $^{\times}$ introduces a tensor, dual to the given vector \cite{fedorov1976}; in components, $(\mathbf{z}^{\times})_{ij} = \varepsilon_{ikj} z^{k}$, where the indices $i,j,k$ run from 1 to 3 (summation over $k$ is implied), and $\varepsilon_{ijk}$ is the Levi-Civita tensor. Such a notation is useful since it allows one to write the cross product of two vectors as a product of a dual tensor and a vector: $\mathbf{u}_1^{\times}(\mathbf{u}_2) \equiv \mathbf{u}_1^{\times} \mathbf{u}_2 = \mathbf{u}_1 \times \mathbf{u}_2.$ Throughout this work, we will regularly employ the following useful identity related to dual tensors: $-\mathbf{z}^{\times} \mathbf{z}^{\times} = \mathbbm{1} - \mathbf{z} \otimes \mathbf{z} \equiv P_t$, where $\otimes$ represents the tensor (Kronecker) product of two vectors, $\mathbbm{1}$ is the identity operator in 3-dimensional space, and $P_t$ is the projector onto the plane orthogonal to $\mathbf{z}$, acting as the identity operator in that plane. A tensor product of two vectors $\mathbf{u}_1$ and $\mathbf{u}_2$ is sometimes called a dyad, and it acts on an arbitrary vector $\mathbf{v}$ of suitable dimension as $(\mathbf{u}_1 \otimes \mathbf{u}_2) \mathbf{v} = \mathbf{u}_1 (\mathbf{u}_2 \cdot \mathbf{v})$. A dyad of the form $\mathbf{u} \otimes \mathbf{u}$ with $\mathbf{u}^2 = 1$ projects the vector it acts upon onto the direction of $\mathbf{u}$.

With an eye towards the case of multiple layers, we would like to turn Eq. (\ref{FourierAmpsEqs}) into a system of equations for the tangential components of the EM fields, $\tilde{\mathbf{H}}_t \equiv P_t \tilde{\mathbf{H}}$ and $\tilde{\mathbf{E}}_t \equiv P_t \tilde{\mathbf{E}}$, since they are continuous across the interface of two media provided surface currents are absent, which we assume is the case. In fact, it is even more convenient to use another tangential vector $\mathbf{z} \times \tilde{\mathbf{E}}$ instead of $\tilde{\mathbf{E}}_{t}$; the reason for that will become clear later, when we introduce the surface impedance tensor. For brevity, we will combine these fields into a single vector $\tilde{\mathbf{W}} \equiv ( \tilde{\mathbf{H}}_t, \mathbf{z} \times \tilde{\mathbf{E}} )^{T}$. In a Cartesian basis with the third axis aligned with $\mathbf{z}$, $\tilde{\mathbf{H}}_t = (\tilde{H}_1, \tilde{H}_2, 0)^{T}$ and $\mathbf{z} \times \tilde{\mathbf{E}} = (-\tilde{E}_2, \tilde{E}_1 , 0)^{T}$.

After a series of linear manipulations with Eq. (\ref{FourierAmpsEqs}), which are presented in detail Appendix \ref{app:1}, we arrive at the following equation:
\begin{equation} \label{basicEq}
	\frac{\mathrm{d} \tilde{\mathbf{W}}}{\mathrm{d} z} = i k_0 M \tilde{\mathbf{W}} + \tilde{\mathbf{U}},
\end{equation} 
where the explicit form of the $M$ operator is given in Appendix \ref{app:3}, and $\tilde{\mathbf{U}}$ is related to the transformed electric current via (\ref{U}). Eq. (\ref{basicEq}) is the desired equation governing the evolution of $\tilde{\mathbf{W}}$ along the $z$ axis. From the mathematical point of view, it is an inhomogeneous ordinary differential equation of the first order. The solution to such an equation is the sum of a general solution to the homogeneous equation (that is, when $\tilde{\mathbf{U}} = 0$) and a particular solution to the inhomogeneous equation \cite{hassani2013}. Since for a homogeneous medium $M$ is a constant operator, the general solution to the homogeneous equation is $\exp[i k_0 M (z - z_0)] \tilde{\mathbf{W}}(z_0) \equiv \Omega_{z_0}^{z} \tilde{\mathbf{W}}(z_0)$, where the constant vector $\tilde{\mathbf{W}} (z_0)$ is the value of $\tilde{\mathbf{W}}$ at some arbitrary point $z_0$. Then, the solution to Eq. (\ref{basicEq}) reads as (\cite{stone2009, hassani2013},
\footnote{The inhomogeneous term can be obtained either by the variation of parameter, or by using the Green's function of Eq. (\ref{basicEq}), which has the form $g(z, z') = \theta(z - z') ~ \Omega^{z}_{z'}$, $\theta$ being the Heaviside step function: $\int_{z_0}^{z} \Omega_{z'}^{z} \tilde{\mathbf{U}}(z') \mathrm{d} z' = \int g(z, z') \tilde{\mathbf{U}}(z') \mathrm{d} z'$. The usual definition of the Green's function $g(z, z')$ is $L g(z,, z') = \delta(z - z')$, where $L$ is the differential operator of the equation under study, and $\delta$ is the delta function. But in the case of Eq. (\ref{basicEq}), we have $L g(z, z') = \delta(z - z') \Omega_{z'}^{z}$. However, this is not a problem since all properties of the delta function are preserved if $\delta$ is multiplied by a function (or an operator) which equals to unity at $z = z'$.})
\begin{equation} \label{homogeneousSolution}
	\tilde{\mathbf{W}}(z) 
	  = \Omega_{z_0}^{z} \tilde{\mathbf{W}}(z_0) + \int \limits_{z_0}^{z} \Omega_{z'}^{z} \tilde{\mathbf{U}}(z') \mathrm{d} z'.
\end{equation}
In the absence of the source term, $\Omega_{z_0}^z$ would relate the values of $\tilde{\mathbf{W}}$ at two points along the $\mathbf{z}$ axis; hence, we will call $\Omega_{z_0}^{z}$ the evolution operator.  Note that $\Omega_z^z$ is equal to the six-dimensional identity operator.

\subsection{Applying interface conditions}

\begin{figure*}
\includegraphics[scale=0.6]{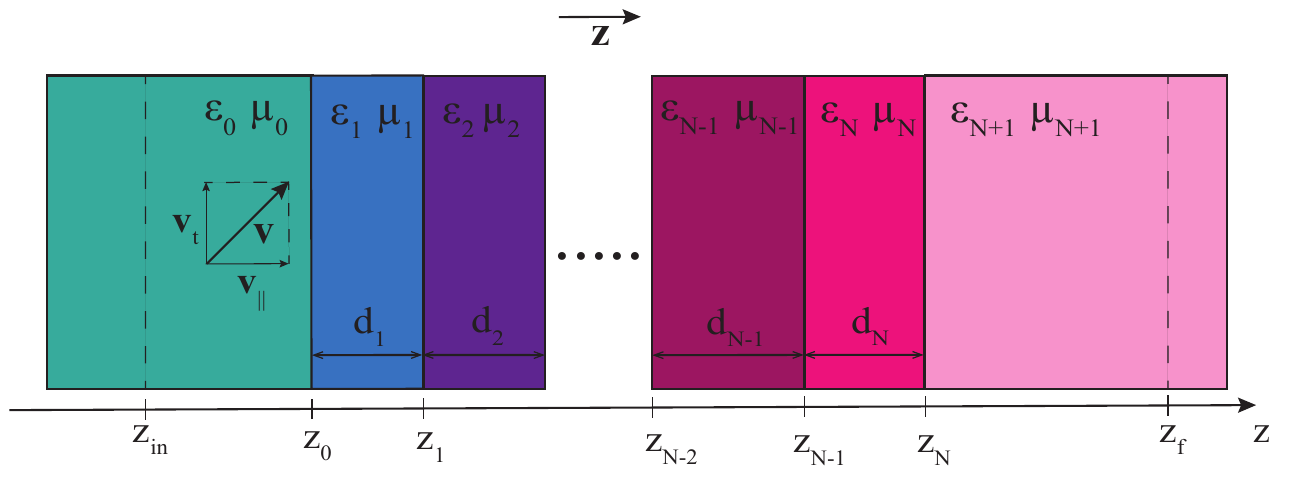}
\caption{\label{Fig1} The layered medium under study. There are $N$ layers, each one is orthogonal to $\mathbf{z}$, has an infinite extent in the tangential plane and has material parameters $\varepsilon_i, \mu_i$ and thickness $d_i$. The layered medium is bounded by semi-infinite media from both sides. The interfaces between layers are located at points $z_0, z_1, \dots, z_N$, and all the currents that generate the fields lie between $z_{\text{in}}$ and $z_{\text{f}}$. An arbitrary vector $\mathbf{v}$ can be split into two components $\mathbf{v}_t$ and $\mathbf{v}_{\parallel}$, which are orthogonal and parallel to $\mathbf{z}$ respectively.}
\end{figure*}

Now we consider a system of $N$ planar layers with material parameters $\varepsilon_i$ and $\mu_i$ and thicknesses $d_i$, as depicted in Fig. \ref{Fig1}.
We assume that each layer is of infinite extent in the plane orthogonal to $\mathbf{z}$, so that the reflections from side interfaces can be neglected. In practice, this means that the tangential dimensions of each layer must be much larger than $k_0 d_i$. The first layer starts at $z = z_0$, the last layer ends at $z = z_N$, and the whole system is bounded by two semi-infinite media with material parameters $\varepsilon_0, \mu_0$ on one side and $\varepsilon_{N+1}, \mu_{N+1}$ on the other side.

Any physically viable current density $\tilde{\mathbf{j}}$ occupies only a finite volume of space. So, without loss of generality, we assume that $\tilde{\mathbf{j}}$ and, hence, $\tilde{\mathbf{U}}$ are nonzero only inside some finite interval of the $z$ axis, which we denote as $(z_{\text{in}}, z_{\text{f}})$, where $z_{\text{in}} < z_0$ and $z_{\text{f}} > z_{N}$ (we elaborate on this point in the Discussion).

Provided that the surface currents are absent, the interface conditions are
\begin{equation}
    \begin{cases}
        \mathbf{z} \times (\mathbf{E}_2 - \mathbf{E}_1) = 0, \\
        \mathbf{z} \times (\mathbf{H}_2 - \mathbf{H}_1) = 0,
    \end{cases}
\end{equation}
where indices $1$ and $2$ label fields from either side of the interface. These are simply the statements that tangential components of the electric and magnetic fields are continuous across any interface. Since the Fourier transform is linear, one can easily show that the interface conditions above imply the continuity of $\tilde{\mathbf{H}}_t$ and $\mathbf{z} \times \tilde{\mathbf{E}}$ across interfaces and, hence, the continuity of $\tilde{\mathbf{W}}$. If we denote the field inside the $i$-th layer at the $j$-th interface as $\tilde{\mathbf{W}}^{(i)}_{z_j}$ (for brevity, from now on we will denote the $z$-dependence of vectors with subscripts), then the interface conditions for $\tilde{\mathbf{W}}$ take the following form:
\begin{eqnarray} \label{interfaceConditions}
\tilde{\mathbf{W}}^{(1)}_{z_0} = \tilde{\mathbf{W}}^{(0)}_{z_0}, ~ \tilde{\mathbf{W}}^{(2)}_{z_1} = \tilde{\mathbf{W}}^{(1)}_{z_1}, ~ \dots, \tilde{\mathbf{W}}^{(N+1)}_{z_N} = \tilde{\mathbf{W}}^{(N)}_{z_N}.
\end{eqnarray}
Thus, $\tilde{\mathbf{W}}$ is a continuous vector function on any interval of the $z$ axis. Substitution of the homogeneous solution (\ref{homogeneousSolution}) into the interface conditions (\ref{interfaceConditions}) yields an expression, relating the fields at the points $z_{\text{in}}$ and $z_{\text{f}}$:
\begin{eqnarray}\label{interfaceConnection}
    &&\tilde{\mathbf{W}}^{(N+1)}_{z_{\text{f}}} = \Omega_{z_{\text{in}}}^{z_{\text{f}}} \tilde{\mathbf{W}}^{(0)}_{z_{\text{in}}} + \int \limits_{z_{\text{in}}}^{z_{\text{f}}} \Omega_{z'}^{z_{\text{f}}} \tilde{\mathbf{U}}_{z'} \mathrm{d} z', \\
    &&\Omega_{z_{\text{in}}}^{z_{\text{f}}} \equiv \Omega_{z_{N}}^{z_{\text{f}}} \left( \prod_{k = 1}^{N} \Omega_{z_{k-1}}^{z_k} \right) \Omega_{z_{\text{in}}}^{z_0}. \nonumber
\end{eqnarray}

After obtaining $\tilde{\mathbf{W}}^{(0)}_{z_{\text{in}}}$ from (\ref{interfaceConnection}) (see Appendix \ref{app:2} for details), it can be exploited to determine the tangential fields at \textit{any} point inside some $j$-th layer, by the virtue of Eq. (\ref{homogeneousSolution}) and the continuity of $\tilde{\mathbf{W}}$ \footnote{We could have expressed $\tilde{\mathbf{W}}_{z_{\text{f}}}^{(N+1)}$ instead of $\tilde{\mathbf{W}}^{(0)}_{z_{\text{in}}}$ and use it in the expressions to follow, which would result in a slightly different form of the DGF, but it would be just as good for practical purposes.}. Thus, we write
\begin{equation}
        \tilde{\mathbf{W}}_{z}^{(j)} = \Omega_{z_{\text{in}}}^{z} \tilde{\mathbf{W}}^{(0)}_{z_{\text{in}}} + \int \limits_{z_{\text{in}}}^{z} \Omega_{z'}^{z} \tilde{\mathbf{U}}_{z'} \mathrm{d} z',
\end{equation}
where $z$ is now an arbitrary point, which may also lie outside of $(z_{\text{in}}, z_{\text{f}})$.

Next, for the known $\tilde{\mathbf{W}}_{z}^{(j)}$, we use Eq. (\ref{realWorld}) to obtain $\mathbf{H}$ and $\mathbf{E}$ at any point $\mathbf{r}$ of the physical space. After that, we compare the resulting fields with the definition of the DGF (\ref{DGFdef}), to finally obtain the dyadic Green's function for both electric and magnetic fields:
\begin{widetext}
\begin{equation} \label{DGFfin}
    \hat{G}^{(j)}(\mathbf{r}, \mathbf{r}') = \dfrac{k_0^2}{\pi c} \int \tilde{G}^{(j)}(\mathbf{b}; z, z') e^{i k_0 \mathbf{b} (\mathbf{r}_{t} - \mathbf{r}_{t}')} ~ \mathrm{d}^2 \mathbf{b} - \frac{4 \pi i}{c k_0 \varepsilon_{\mathbf{z}}^{(j)}}
		\begin{pmatrix}
			0 \\
			\mathbf{z} \otimes \mathbf{z}
		\end{pmatrix} \delta(\mathbf{r} - \mathbf{r}'),
\end{equation}
where
\begin{gather} \label{AfterDGF}
    \tilde{G}^{(j)}(\mathbf{b}; z, z') \equiv V^{(j)} P_2 \left( \Omega_{z_{\text{in}}}^{z} S \Omega_{z'}^{z_{\text{f}}} + \theta(z - z') \Omega_{z'}^{z} \right) Q, \\
    S \equiv -\begin{pmatrix}
			P_t \\
			-\gamma^{(0)}
		\end{pmatrix}
		\left[
			\begin{pmatrix}
				\gamma^{(N+1)}, & -P_t 
			\end{pmatrix} \Omega_{z_{\text{in}}}^{z_{\text{f}}}
		\begin{pmatrix}
			P_t \\
			-\gamma^{(0)}
		\end{pmatrix}
		\right]^{-}
			\begin{pmatrix}
				\gamma^{(N+1)}, & -P_t 
			\end{pmatrix}, \quad Q \equiv
			\frac{1}{\varepsilon_{\mathbf{z}}}
	\begin{pmatrix}
		\mathbf{z}^{\times} \varepsilon (\mathbf{z} \otimes \mathbf{z}) \\
		-\mathbf{b}^{\times} (\mathbf{z} \otimes \mathbf{z})
	\end{pmatrix} -
	\begin{pmatrix}
		\mathbf{z}^{\times} \\
		0
	\end{pmatrix}  .
\end{gather}
\end{widetext}

We want to point out that since $\mathbf{H}$ and $\mathbf{E}$ are, in general, discontinuous across interfaces, the DGF is also a piecewise-continuous tensor function. So, it has a label $(j)$ attached to it, which reminds us that $\hat{G}^{(j)}$ should be used to obtain fields only inside the $j$-th layer. Also, as was said earlier, we can calculate both $\mathbf{E}$ and $\mathbf{H}$ using the obtained DGF, rather than $\mathbf{E}$ alone. If we write $\hat{G}^{(j)} = (\hat{G}_H^{(j)}, ~ \hat{G}_E^{(j)})^{T}$, then it is $\hat{G}_E$ which corresponds to the “usual” electric DGF of dimension $3 \times 3$. However, in order not to multiply the entities, we shall still refer to the whole of $\hat{G}$ as DGF.

\subsection{Discussion}

Let us spell out how to interpret the obtained expression. The DGF $\hat{G}^{(j)}$ in (\ref{DGFfin}) is an operator, represented in some fixed basis as a $6 \times 3$ matrix, which should be convoluted with a $3$-dimensional current density vector $\mathbf{j}$ in order to obtain a $6 \times 1$ vector made up of magnetic and electric fields:
\begin{equation}
    \begin{pmatrix}
		\mathbf{H}^{(j)} (\mathbf{r}) \\
		\mathbf{E}^{(j)} (\mathbf{r}) 
	\end{pmatrix} = \int\limits_{V} \hat{G}^{(j)}(\mathbf{r}, \mathbf{r}') \mathbf{j}(\mathbf{r}') \mathrm{d}^3 \mathbf{r}',
\end{equation}
where the integration is performed over $(z_{\text{in}}, z_{\text{f}}) \times \mathbb{R}^2$ or, if the currents are defined on a smaller subset of $\mathbb{R}^3$, the integration is over that subset. The resulting field distribution is only valid within the $j$-th layer since, unlike their tangential components, the full $\mathbf{H}$ and $\mathbf{E}$ fields are discontinuous across interfaces. All the quantities which have the $(j)$ label should be taken as constants when integrating over $\mathbf{r}'$, while the unlabeled operators like $Q$ and $\Omega_{z'}^{z}$ should be regarded as piecewise-continuous functions with respect to $z'$. Remember that if one wishes to obtain the experimentally observable electric and magnetic fields, they should take the real parts of $\mathbf{H}^{(j)}$ and $\mathbf{E}^{(j)}$.

The introduction of the interval $(z_{\text{in}}, z_{\text{f}})$, which contains all the electric currents within itself, may have appeared artificial. Indeed, except for the requirement that $\tilde{\mathbf{j}}$ should vanish outside of this interval, the choice of $z_{\text{in}}$ and $z_{\text{f}}$ is completely arbitrary. This interval was introduced for the following reason. The physical intuition tells us that a current, located at some point $z$, should, in general, contribute to the value of the EM field at any other point in space. Thus, whatever is the final solution to (\ref{basicEq}) for the entire layered medium, we expect it to contain the contributions from all the currents in the system. By investigating (\ref{homogeneousSolution}), we conclude that it would be so if the integration in the second term is performed over all the values of $z$ where $\tilde{\mathbf{U}}$ does not vanish. However, such is the nature of first-order differential equations that in (\ref{interfaceConnection}) we could have related the values of $\tilde{\mathbf{W}}$ at $z_{\text{in}}$ and some arbitrary point $z_{\text{f}}' < z_{\text{f}}$, then express $\tilde{\mathbf{W}}_{z_\text{in}}$ and obtain a formula similar to Eq. (\ref{W0Formula}), which would still be a solution to (\ref{basicEq}), but would not correspond to the problem at hand, because then, in the expression for $\tilde{\mathbf{W}}_{z_\text{in}}$, not all of the electric currents would be accounted for. In other words, it \textit{would} be a solution to (\ref{basicEq}), but it \textit{would not} be the one we need. So, in order to make the solution physical, we have assumed the existence of such an interval of the $z$ axis that there are no electric currents outside of it. This assumption is in line with typical experimental setups, when the currents are either due to the emitters located just outside the layered media, or within some of its layers (a model of a light-emitting diode). In principle, one could take the limit $(z_{\text{in}}, z_{\text{f}}) \to (-\infty, \infty)$, but we believe it to be unnecessary for any practical purposes.

Our definition (\ref{DGFdef}) of the DGF ostensibly differs from that of, e.g., \cite{novotny2012}, where it is defined via the electric field produced by an oscillation dipole: In that case, $\mathbf{j} = - i \omega \mathbf{d} \delta(\mathbf{r} - \mathbf{r}_{d})$, and (in Gaussian units) $\mathbf{E} = 4 \pi k_0^2 \hat{G}_{E}(\mathbf{r}, \mathbf{r}_{d}) \mathbf{d}$. However, this latter definition only differs from ours by a constant factor of $4 \pi i k_0 / c$, since, using the definition (\ref{DGFdef}), the field due to an oscillating dipole would be $- i \omega \hat{G}_E \mathbf{d}$. So, if one wishes to be in concordance with the more common definition, one should divide the expression (\ref{DGFfin}) by $4 \pi i k_0 / c$. Then, the coefficient in front of the $\delta$-function term will become real (for non-absorbing media), and it becomes manifest that the imaginary part of $\hat{G}$ does not have a singularity at $\mathbf{r} = \mathbf{r}'$ — a fact which is well-known for simple media, but, in general, is difficult to prove.

The problem we have solved is a stationary one, which means that a sufficiently long time has passed since the currents have been turned on, so that all the possible relaxation processes have ended, and we can deal with the time-harmonic distribution of fields. What period of time can be regarded as “sufficiently long”? The natural time scale for this problem is $d / c$, where $d$ is the total thickness of the layered system, so it should be $t \gg d / c$ in order for the obtained results to hold. But still, this period of time is tiny compared to usual macroscopic time scales, due to the largeness of the speed of light.

We have examined anisotropic layers, which are defined by the constitutive relations $\mathbf{D} = \varepsilon \mathbf{E}$ and $\mathbf{B} = \mu \mathbf{H}$, where $\varepsilon$ and $\mu$ are tensors. However, there are also so-called bi-anisotropic media, characterized by more general constitutive relations
\begin{equation}
    \begin{cases}
        \mathbf{D} = \varepsilon \mathbf{E} + \alpha \mathbf{H}, \\
        \mathbf{B} = \beta \mathbf{E} + \mu \mathbf{H},
    \end{cases}
\end{equation}
where $\alpha$ and $\beta$ are also tensors. Such media are encountered, e.g., when modeling metamaterials which encapsulate the equations of axion electrodynamics \cite{gorlach2023}. Even though we do not present the corresponding calculations, the DGF for such media may also be found using the methods employed in this paper: The only difference would be in the form of the $M, \gamma$ and $V$ operators (their form for bi-anisotropic media can be found in \cite{Borzdov97}) and of the inhomogeneous term $\tilde{\mathbf{U}}$. 

Finally, it is interesting to note a couple of intersections between our derivation of the DGF and other areas of physics. First, we were able to find the distribution of the EM field at every point in space using only its value in the plane $ z = z_{\text{in}} $. Thus, we have incidentally confirmed the mathematical consistency of hologrpahy, which tells that a three-dimensional scene can be reconstructed using a diffraction pattern it produces on a two-dimensional plate \cite{born1970}. Second, there is a direct mathematical analogy between Eq. (\ref{basicEq}) and the Schrödinger's equation $ i \partial | \psi \rangle / \partial t = \hat{H} | \psi \rangle$. The operator $- k_0 M$ plays the role of the Hamiltonian $\hat{H}$: whereas the latter governs the evolution of the state vector $| \psi \rangle$ in time, the former governs the spatial evolution of $\tilde{\mathbf{W}}$ along the $z$ axis. Thus, by analogy, $- k_0 M$ may be regarded as the generator of infinitesimal translations along the $z$ axis. If $\hat{H}$ does not explicitly depend on time, then $| \psi (t) \rangle = e^{- i \hat{H} (t - t_0)} | \psi (t_0) \rangle \equiv \hat{U}(t, t_0) | \psi \rangle$, similar to $\tilde{\mathbf{W}}(z) = e^{i k_0 M (z - z_0)} \tilde{\mathbf{W}}(z_0) \equiv \Omega_{z_0}^z \tilde{\mathbf{W}}(z_0)$, which holds when $\varepsilon$ and $\mu$, and hence $M$, do not depend on $z$ (and if $\tilde{\mathbf{U}} = 0$). If $\hat{H}$ \textit{does} depend on time, one can approximate $\hat{U}(t, t_0)$ by splitting the interval $t - t_0$ into $N$ segments and regard $\hat{H}$ to be constant on each of them. Then $\hat{U}(t, t_0) \approx \prod_{i = 1}^{N} \exp[- i \hat{H}_{t_i} (t_i - t_{i-1})]$. In electrodynamics, this corresponds to the problem we have investigated: material parameters of the medium \textit{do} depend on $z$, but are constant within each of the $N$ layers. In general, whenever $\hat{H}$ depends explicitly on $t$ and does not commute with itself at different moments of time, $\hat{U}(t, t_0)$ can be written as a time-ordered exponential, also called the Dyson series \cite{dyson1949}, and a similar formula can be written for $\Omega_{z_0}^{z}$ (“$z$-ordered exponential”), which corresponds to a continuously inhomogeneous medium \cite{Borzdov97}.

\section{\label{sec:3} Applications}

Now that we have the expression for the DGF, let us provide a couple of examples of its use. 

The generality of the obtained DGF comes with a downside: It is hard to extract much analytic results from (\ref{DGFfin}), mainly because of the pseudoinverse operation in the expression for $S$. However, in all but the simplest cases, for which the closed-form expressions for the DGF are already known, one resorts to a numerical evaluation of the DGF. Thus, it would be instructive to perform several numerical calculations and point out useful tips on how to effectively use the obtained expression in such calculations.

In order to perform numerical calculations, we have to choose some specific basis $(\mathbf{e}_1, \mathbf{e}_2, \mathbf{e}_3)$. Throughout the rest of this chapter, we choose the $\mathbf{e}_{i}$ vectors such that $\mathbf{e}_3 = \mathbf{z}$ and the whole triplet is orthonormal and right-handed. 

\begin{figure}[ht]
\includegraphics[scale=0.3]{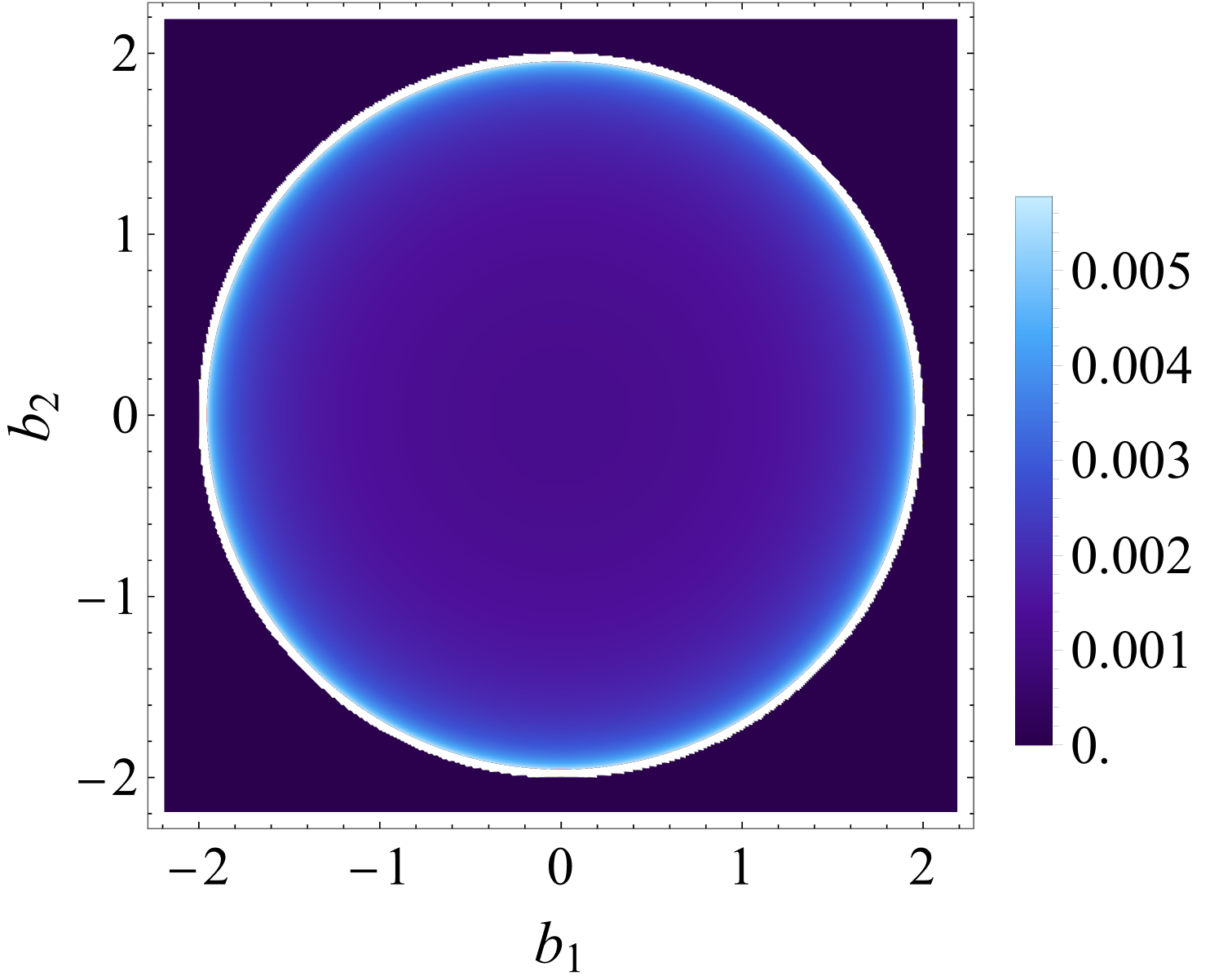}
\caption{\label{Fig2} A typical form of the integrand in the expression for the LDOS (\ref{rhoviaDGF}). The integrand is rotationally symmetric whenever there are no preferred directions in the tangential plane.}
\end{figure}

\begin{figure}[ht]
\includegraphics[scale=0.45]{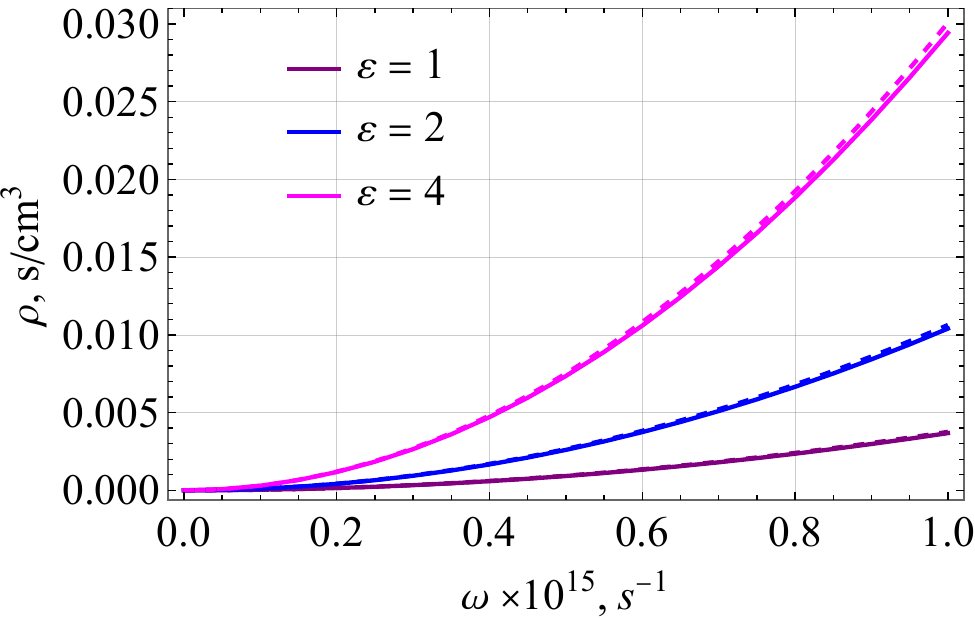}
\caption{\label{Fig3} Comparison of the analytic expression (\ref{rhoAnalytic}) for the LDOS (dashed lines) with numerical evaluations of (\ref{rhoviaDGF}) (solid lines) for different permittivities.}
\end{figure}

First, consider a single homogeneous isotropic medium of permittivity $\varepsilon$. We want to calculate the local density of states (LDOS), the analytic formula for which is \cite{barnes2020}
\begin{equation}\label{rhoAnalytic}
    \rho(\mathbf{r}, k_0) = \dfrac{\varepsilon^{3/2} k_0^2}{\pi^2 c}.
\end{equation}
It can also be expressed via the electric DGF as
\begin{equation}
    \rho(\mathbf{r}, k_0) = \dfrac{2 k_0 \varepsilon}{\pi c} \operatorname{Tr} \operatorname{Im} \hat{G}_E (\mathbf{r}, \mathbf{r}, k_0).
\end{equation}
Using Eq. (\ref{DGFfin}) divided by $4 \pi i k_0 / c$ (for the reasons explained in the previous section), the last equation boils down to
\begin{equation}\label{rhoviaDGF}
    \rho(\mathbf{r}, k_0) = -\dfrac{\varepsilon k_0^2}{2 \pi^3 c} \int \operatorname{Tr} \operatorname{Re} \tilde{G}_E (\mathbf{b}; z, z) \mathrm{d}^2 \mathbf{b}.
\end{equation}
Now we have to evaluate the integral over the tangential part of the dimensionless wavevector. In general, evaluating a two-dimensional integral of a product of complicated operators can be numerically demanding, but if we examine the integrand in the last equation prior to performing calculations, we can save ourselves some work. A typical behavior of the integrand $\operatorname{Tr} \operatorname{Re} \tilde{G}_E (\mathbf{b}; z, z)$ is presented in Fig. \ref{Fig2}: It is nonzero inside the disk $\mathbf{b}^2 < \varepsilon$, undefined on the circle $\mathbf{b}^2 = \varepsilon$ (see Appendix \ref{app:3}) and almost instantaneously vanishes outside this circle.
Owing to the rotational symmetry of the integrand, instead of evaluating the full two-dimensional integral, we can instead use the well-known formula for the volume of solids of revolution $V = 2\pi \int_{a_1}^{a_2} x f(x) dx$ in order to evaluate Eq. (\ref{rhoviaDGF}). In that case, $a_1 = 0$, $a_2 = \sqrt{\varepsilon}$, $x = b_1$, and $f(x) = \operatorname{Tr} \operatorname{Re} \tilde{G}_E (\{b_1, 0, 0\}; z, z)$. One should also be cautious when constructing the integration grid: As can be seen in Fig. \ref{Fig2}, the integrand is almost constant near the center of the disk, but rises quickly to some considerable yet finite value near the circle $\mathbf{b}^2 = \varepsilon$. Thus, using a homogeneous grid may be ineffective, since such a grid may miss the contributions to the integral that come from near the boundary. Instead, it is expedient to use an inhomogeneous grid, for which the points become denser near the edge of the integration interval. The results of performing numerical evaluation of Eq. (\ref{rhoviaDGF}) are presented in Fig. \ref{Fig3}. 
We have used an integration grid with $200$ points on the interval $(0, 1)$, weighted by $\sqrt{\varepsilon}(1 - e^{-10 b_1})$, and the results are in good concordance with the analytic expression. In general, whenever one wants to perform calculations for some complicated multilayered medium, it is instructive first to perform these calculations for some simple medium, for which the closed-form analytic expression for the DGF is known, and compare the numerical and analytic results. This would then serve as the “calibration” of one's numerical scheme.

Next, we consider two layered isotropic media (everywhere below, we take the angular frequency $\omega$ to be $10^{15} ~ \text{rad s}^{-1}$, and $\lambda = 2\pi c / \omega$). The first one consists of a single layer with permittivity $\varepsilon = 4$ and thickness $d = \lambda$, and the second one consists of three layers: two outer layers have permittivity $\varepsilon_1 = 4$ and thickness $d_1 = \lambda$, while the inner layer has $\varepsilon_2 = 2$ and $d_2 = \lambda/2$. In both cases, the layers are surrounded by vacuum from either sides. The results for the LDOS enhancement compared to its vacuum value $\rho_0$, as a function of distance for such media are presented in Fig. \ref{Fig3}.
\begin{figure}[ht]
\includegraphics[scale=0.45]{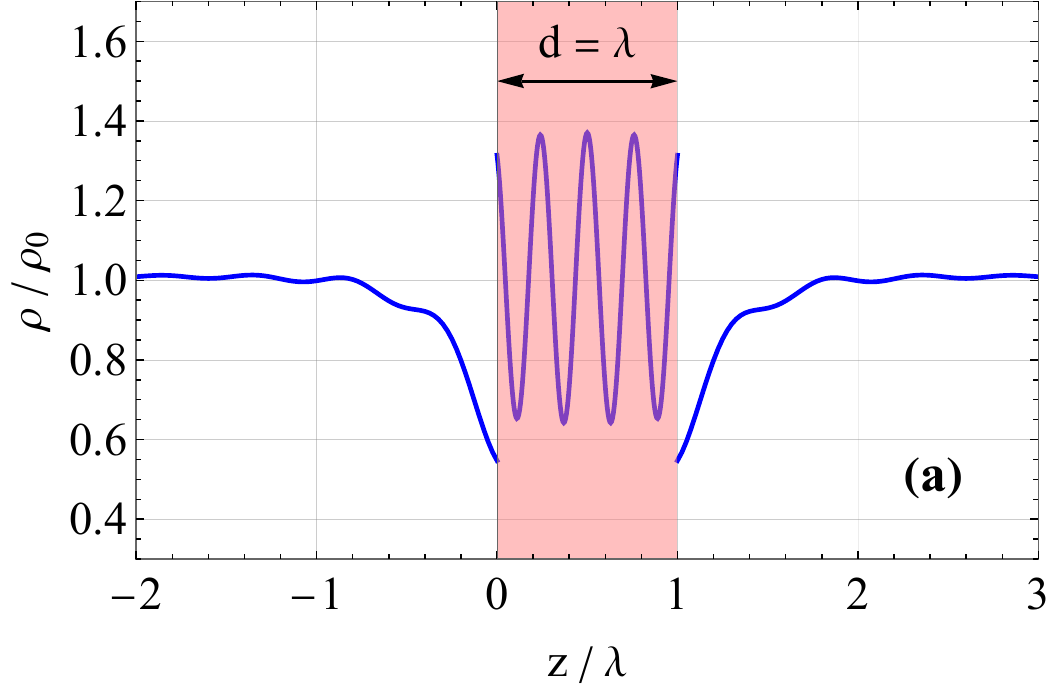}
\includegraphics[scale=0.45]{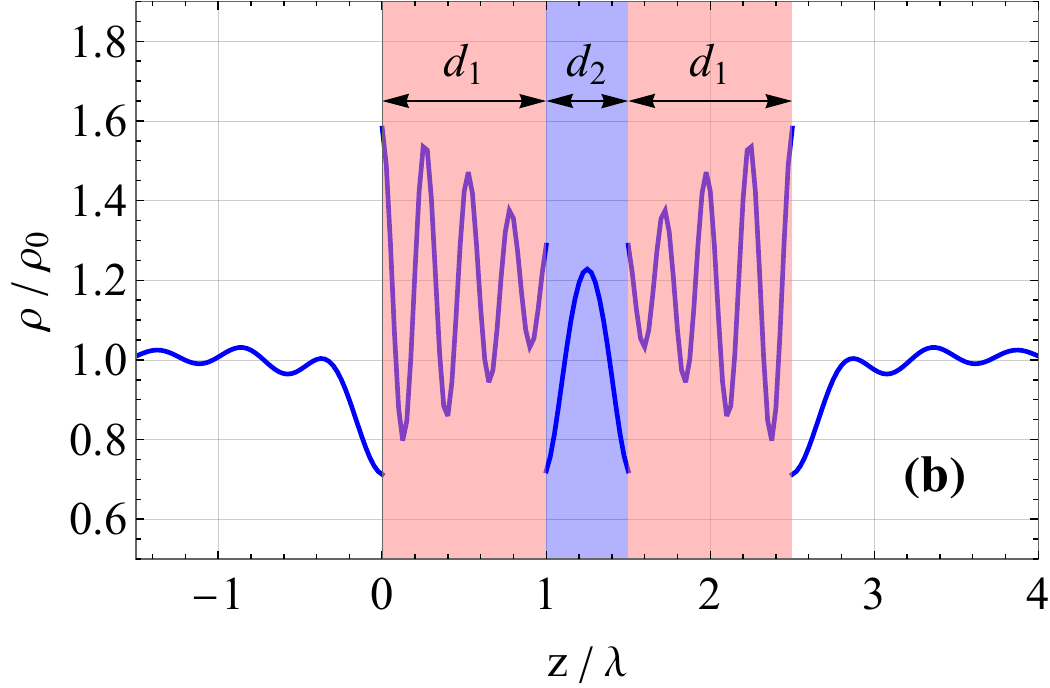}
\caption{\label{Fig4} LDOS enhancement compared to its vacuum value as a function of distance for (a) a single layer with $\varepsilon = 4$ and thickness $d = \lambda$ (red area) surrounded by vacuum; (b) three layers, two of which have $\varepsilon_1 = 4$ and $d_1 = \lambda$ (red areas) and one with $\varepsilon_2 = 2$ and $d_2 = \lambda/2$ (blue area).}
\end{figure}

We have used the same integration scheme as before, only now the radius of the integration disk is $\sqrt{\varepsilon_2}$. In general, whenever the medium consists of $N$ isotropic non-magnetic layers, the radius of the integration disk is $\sqrt{\min(\varepsilon_1, \varepsilon_2, \dots, \varepsilon_N)}$.

Finally, we consider an anisotropic layered medium, with the anisotropy pertinent to gyrotropic media \cite{fedorov1976}, for which the permittivity tensor is
\begin{equation}\label{epsGTensor}
    \varepsilon_{\text{g}} = \begin{pmatrix}
        \varepsilon_0 & i g & 0 \\
        -ig & \varepsilon_0 & 0 \\
        0 & 0 & \varepsilon_0
    \end{pmatrix}.
\end{equation}
Gyrotropy in a medium can either be induced by external fields, as in magnetooptics, or it can be an intrinsic property of the material, as in recently discovered Weyl semimetals, wherein the gyrotropy is caused by the so-called axion term \cite{Armitage2018,Guo2023}. 

Since the permittivity tensor (\ref{epsGTensor}) can be represented as
\begin{equation}
\varepsilon_g = 
\begin{pmatrix}
        \varepsilon_0 & i g & 0 \\
        -ig & \varepsilon_0 & 0 \\
        0 & 0 & 0
    \end{pmatrix} + \varepsilon_0 \mathbf{z} \otimes \mathbf{z}, 
\end{equation}
where the first term on the right-hand side is a Hermitian matrix, there is no preferred direction in the tangential plane, and we can still employ the rotational symmetry of the integrand in (\ref{rhoviaDGF}) and use the same numerical scheme as in the previous examples. If, however, there \textit{was} a preferred direction, there would be no such symmetry, and the whole two-dimensional integral would have to be evaluated.

\begin{figure}[ht]
\includegraphics[scale=0.42]{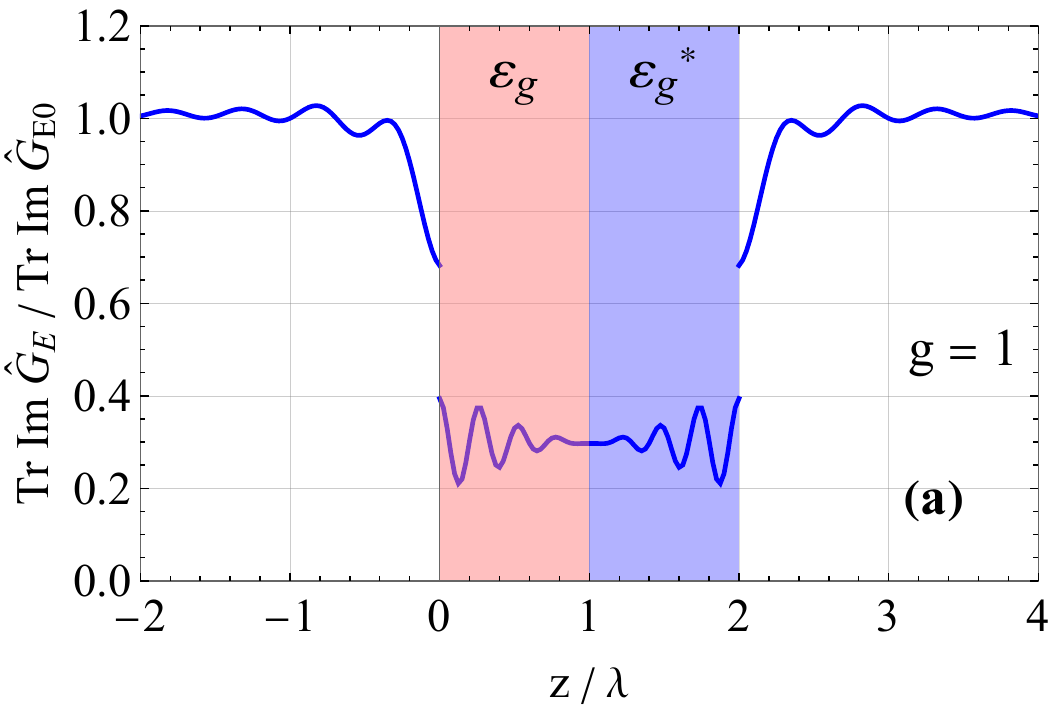}
\includegraphics[scale=0.42]{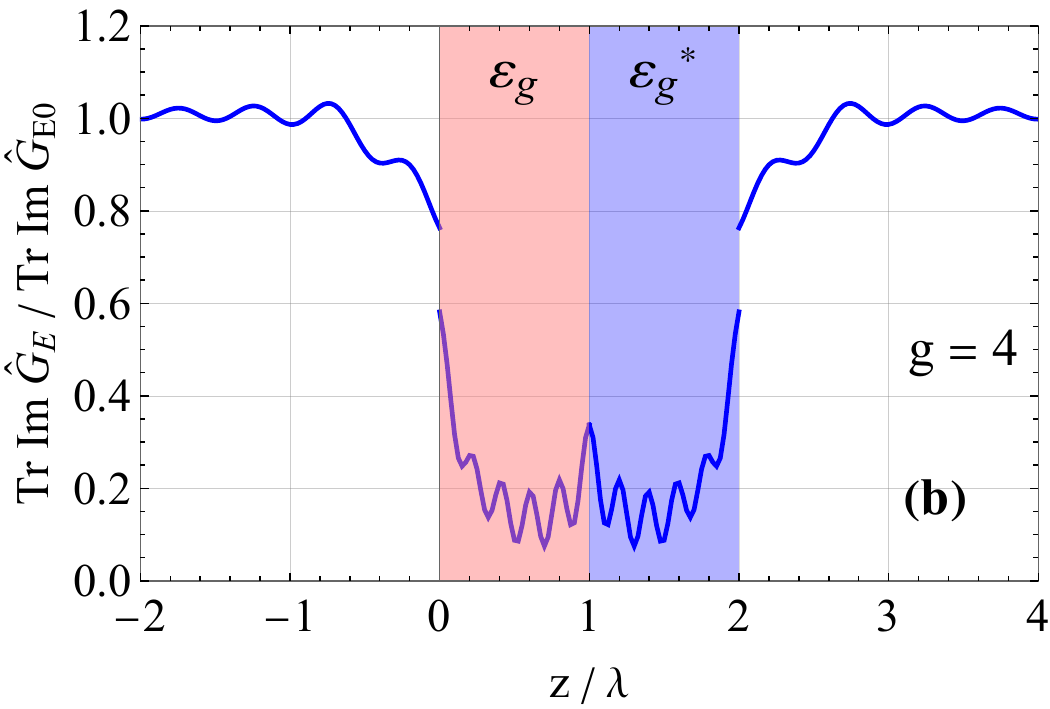}
\caption{\label{Fig5}
$\operatorname{Tr} \operatorname{Im} \hat{G}_{E}$ as a function of distance for the bi-layered gyrotropic medium, the permittivity tensors of the two layers being complex-conjugate of one another, with parameters $\varepsilon_0 = 4$ and (a) $g = 1$; (b) $g = 4$.}
\end{figure}

We calculate $\operatorname{Tr} \operatorname{Im} \hat{G}_{E}$, a quantity to which the LDOS is proportional to, for a bi-layered medium composed of two gyrotropic layers of thicknesses $\lambda$, with the permittivity tensors $\varepsilon_g$ for the first layer and complex-conjugated $\varepsilon_g^*$ for the second layer. We take $\varepsilon_0 = 4$ and consider two values of the $g$ parameter, $g = 1$ and $g = 4$. As in the previous example, we express $\operatorname{Tr} \operatorname{Im} \hat{G}_{E}$ in units of its vacuum value, $\operatorname{Tr} \operatorname{Im} \hat{G}_{E0}$. The results are presented in Fig. \ref{Fig5}.

\section{\label{sec:4}Summary}

We have derived the dyadic Green's function of a generic anisotropic planarly-layered medium. This was achieved by directly solving the Maxwell's equations with electric currents for such a medium and comparing the results with the definition of the electric DGF. The obtained expression (\ref{DGFfin}) is valid for an arbitrary number of layers, each having arbitrary $\varepsilon$ and $\mu$ tensors and thicknesses. The use of the operator approach, i.e., the treatment of vectors and tensors as geometric objects rather than arrays of numbers, has simplified the process of derivation and the form of the resulting expression, in that it made all the calculations clearer and more concise. When using Eq. (\ref{DGFfin}), one has to remember that it was derived based on the definition (\ref{DGFdef}), which differs from a more common one by a constant factor of $4\pi i k_0 / c$.

Though it may be hard to use the resultant expression for the DGF for further analytic computations, it can be readily implemented for numerical calculations. When performing such calculations, one has to use some specific basis, and the freedom in its choice can be utilized, for example, by choosing the basis in which permittivity tensors of layers have the simplest form. Furthermore, by closely inspecting the integrand in (\ref{DGFfin}), one can significantly reduce the amount of calculations: for example, when the medium has no preferred directions in the tangential plane, the rotational symmetry of the integrand in the $(b_1, b_2)$ plane can be exploited to reduce the two-dimensional integral to a one-dimensional one, which greatly decreases the computation time.

Although we focused on the case of planar layers, the general algorithm for obtaining the DGF remains valid for non-planar, e.g., cylindrical or spherical layers, which is briefly discussed in Appendix \ref{app:4}. The main difference lies in the choice of a suitable orthonormal set of functions used to expand the electric and magnetic fields.

\acknowledgments{A.A. and A.N. thank the Belarusian Republican
Foundation for Fundamental Research (Project No. F26RNF-106). D.G. acknowledges the National Natural Science Foundation of China (12174281). This work was also supported by the National Natural Science Foundation of China (12311530763); Natural Science Foundation of Jiangsu Province (BK20221240); Suzhou Basic Research Project (SJC2023003).}

\appendix

\section{\label{app:1} Derivation of the first-order evolution equation}

After dot-multiplying both equations in (\ref{FourierAmpsEqs}) by $\mathbf{z}$, we get
\begin{equation} \label{scalarEqs}
    \begin{cases}
		\tilde{\mathbf{H}} \cdot  \mathbf{a} = (\varepsilon\tilde{\mathbf{E})} \cdot \mathbf{z} + \dfrac{4\pi i}{\omega} \tilde{\mathbf{j}} \cdot \mathbf{z}, \\
		\tilde{\mathbf{E}} \cdot \mathbf{a} = - (\mu\tilde{\mathbf{H}}) \cdot \mathbf{z},
	\end{cases} ~~~ \mathbf{a} \equiv \mathbf{b} \times \mathbf{z}.
\end{equation}
From this system we can express the components of the fields along the $\mathbf{z}$ axis, i.e. $ \tilde{E}_{z} \equiv \mathbf{z} \cdot \tilde{\mathbf{E}}$ and $\tilde{H}_{z} \equiv \mathbf{z} \cdot \tilde{\mathbf{H}}$. To do that, we use the following identity:
\begin{equation}
    (\varepsilon\tilde{\mathbf{E})} \cdot \mathbf{z} = (\varepsilon (P_t + \mathbf{z} \otimes \mathbf{z}) \tilde{\mathbf{E})} \cdot \mathbf{z} = (\varepsilon \tilde{\mathbf{E}}_t) \cdot \mathbf{z} + \tilde{E}_{z} \varepsilon_{\mathbf{z}}
\end{equation}
and, similarly, $(\mu\tilde{\mathbf{H})} \cdot \mathbf{z} = (\mu \tilde{\mathbf{H}}_t) \cdot \mathbf{z} + \tilde{H}_{z} \mu_{\mathbf{z}}$, where $\mu_{\mathbf{z}} \equiv \mathbf{z} \cdot (\mu \mathbf{z})$ and $\varepsilon_{\mathbf{z}} \equiv \mathbf{z} \cdot (\varepsilon \mathbf{z})$. Thus, we get
\begin{equation} \label{zComponents}
    \begin{cases}
        \tilde{H}_{\mathbf{z}} = - \frac{1}{\mu_{\mathbf{z}}} \left( \mathbf{a} \cdot \tilde{\mathbf{E}}_t + \mathbf{z} \cdot (\mu \tilde{\mathbf{H}}_t) \right), \\
        \tilde{E}_{\mathbf{z}} = \frac{1}{\varepsilon_{\mathbf{z}}} \left( \mathbf{a} \cdot \tilde{\mathbf{H}}_t - \mathbf{z} \cdot (\varepsilon \tilde{\mathbf{E}}_t) - \frac{4 \pi i}{\omega} \mathbf{z} \cdot \tilde{\mathbf{j}} \right).
    \end{cases}
\end{equation}
Since any vector $\mathbf{v}$ can be represented as $\mathbf{v} = P_t \mathbf{v}_t + v_{z} \mathbf{z}$, the expressions (\ref{zComponents}) allow us to recover the full Fourier amplitudes using only their tangential components and the source term:
\begin{equation}
\label{connection}
 \begin{split}
	\begin{pmatrix}
		\tilde{\mathbf{H}} \\
		\tilde{\mathbf{E}}
	\end{pmatrix} = 
	&\underbrace{\begin{pmatrix}
		P_t - \frac{1}{\mu_{\mathbf{z}}} (\mathbf{z} \otimes \mathbf{z}) \mu & -\frac{1}{\mu_{\mathbf{z}}} \mathbf{z} \otimes \mathbf{a} \\
		\frac{1}{\varepsilon_{\mathbf{z}}} \mathbf{z} \otimes \mathbf{a} & P_t - \frac{1}{\varepsilon_{\mathbf{z}}} (\mathbf{z} \otimes \mathbf{z}) \varepsilon	\end{pmatrix}}_{V}
	\begin{pmatrix}
		\tilde{\mathbf{H}}_t \\
		\tilde{\mathbf{E}}_t
	\end{pmatrix} \\
    & - \frac{4\pi i}{\omega \varepsilon_{\mathbf{z}}}
	\begin{pmatrix}
		0 \\
		\mathbf{z} \otimes \mathbf{z}
	\end{pmatrix} \tilde{\mathbf{j}}.
 \end{split}
\end{equation} 
The fact that the full $\tilde{\mathbf{H}}$ and $\tilde{\mathbf{E}}$ fields can be determined using only $\tilde{\mathbf{H}}_t$ and $\tilde{\mathbf{E}}_t$ is crucial for our derivation of the DGF.

The system (\ref{FourierAmpsEqs}) can be written in the block matrix form as
\begin{equation} \label{step1}
		\begin{pmatrix}
			\mathbf{z}^{\times} & 0 \\
			0 & \mathbf{z}^{\times}
		\end{pmatrix} \frac{\mathrm{d}}{\mathrm{d} z}
		\begin{pmatrix}
			\tilde{\mathbf{H}} \\
			\tilde{\mathbf{E}}
		\end{pmatrix} = i k_0 
        \underbrace{\begin{pmatrix}
			-\mathbf{b}^{\times} & - \varepsilon \\
			\mu & -\mathbf{b}^{\times}
		\end{pmatrix}}_{N}
		\begin{pmatrix}
			\tilde{\mathbf{H}} \\
			\tilde{\mathbf{E}}
		\end{pmatrix} +
        \begin{pmatrix}
		      \frac{4\pi}{c} \tilde{\mathbf{j}} \\
			0
		\end{pmatrix}.
\end{equation}
Note that
\begin{equation}
		\underbrace{\begin{pmatrix}
			-\mathbf{z}^{\times} & 0 \\
			0 & P_t
		\end{pmatrix}}_{P_1}
		\begin{pmatrix}
			\mathbf{z} \times \tilde{\mathbf{H}} \\
			\mathbf{z} \times \tilde{\mathbf{E}}
		\end{pmatrix} =
		\begin{pmatrix}
			\tilde{\mathbf{H}}_t \\
			\mathbf{z} \times \tilde{\mathbf{E}}
		\end{pmatrix} \equiv \tilde{\mathbf{W}}.
\end{equation}
Multiplying (\ref{step1}) by $P_1$, we get
\begin{equation} \label{intermediate}
	\frac{\mathrm{d} \tilde{\mathbf{W}}}{\mathrm{d} z} = i k_0 P_1 N
	\begin{pmatrix}
		\tilde{\mathbf{H}} \\
		\tilde{\mathbf{E}}
	\end{pmatrix} + P_1
	\begin{pmatrix}
		\frac{4\pi}{c} \tilde{\mathbf{j}} \\
		0
	\end{pmatrix}.
\end{equation}
Also,
\begin{equation} \label{tToW}
		\begin{pmatrix}
			\tilde{\mathbf{H}}_t \\
			\tilde{\mathbf{E}}_t
		\end{pmatrix} =
        \underbrace{\begin{pmatrix}
			P_t & 0 \\
			0 & -\mathbf{z}^{\times}
		\end{pmatrix}}_{P_2}
		\begin{pmatrix}
			\tilde{\mathbf{H}}_t \\
			\mathbf{z} \times \tilde{\mathbf{E}}
		\end{pmatrix} \equiv P_2 \tilde{\mathbf{W}}.
\end{equation}
Upon substituting (\ref{connection}) into (\ref{intermediate}) and then using (\ref{tToW}), we finally arrive at the desired first-order equation (\ref{basicEq}), where
\begin{equation}
	M \equiv P_1 N V P_2,
\end{equation}
\begin{equation} \label{U}
 \tilde{\mathbf{U}} \equiv \frac{4\pi}{c} \left[
    \frac{1}{\varepsilon_{\mathbf{z}}}
	\begin{pmatrix}
		\mathbf{z}^{\times} \varepsilon (\mathbf{z} \otimes \mathbf{z}) \\
		- \mathbf{b}^{\times} (\mathbf{z} \otimes \mathbf{z})
	\end{pmatrix} -
	\begin{pmatrix}
		\mathbf{z}^{\times} \\
		0
	\end{pmatrix} \right] ~ \tilde{\mathbf{j}}.
\end{equation}
Note that the symmetry of the system naturally manifests itself in the source term $\tilde{\mathbf{U}}$, given by Eq. (\ref{U}): if the current $\tilde{\mathbf{j}}$ is orthogonal to $\mathbf{z}$, the axis of symmetry, the first term inside the square brackets vanishes; alternatively, if the current is parallel to $\mathbf{z}$, the second term vanishes.

Once the $\tilde{\mathbf{W}}$ vector is known, one can restore the $\mathbf{H}$ and $\mathbf{E}$ fields in the physical space by invoking Eqs. (\ref{connection}) and (\ref{tToW}):
\begin{equation}
\begin{split} \label{realWorld}
	&\begin{pmatrix}
		\mathbf{H} \\
		\mathbf{E}
	\end{pmatrix} = \int
    \begin{pmatrix}
		\tilde{\mathbf{H}} \\
		\tilde{\mathbf{E}}
	\end{pmatrix} e^{i k_0 \mathbf{b} \mathbf{r}_{t}} ~ \mathrm{d}^2( k_0^2 \mathbf{b} ) \\
    &= \int V P_2 \tilde{\mathbf{W}} e^{i k_0 \mathbf{b} \mathbf{r}_{t}} ~ \mathrm{d}^2( k_0^2 \mathbf{b} ) - \frac{4\pi i}{c k_0 \varepsilon_{\mathbf{z}}}
	\begin{pmatrix}
		0 \\
		\mathbf{z} \otimes \mathbf{z} 
	\end{pmatrix} \mathbf{j}.
\end{split}
\end{equation}

\section{\label{app:2} Solving the evolution equation for a multilayered medium}

In order to solve Eq. (\ref{interfaceConnection}) (it is a system of six equations, two of which are of the trivial form $0 = 0$), i.e. to find $\tilde{\mathbf{W}}^{(0)}_{z_{\text{in}}}$, we introduce the surface impedance tensor $\gamma^{(i)}$, which relates $\mathbf{z} \times \tilde{\mathbf{E}}^{(i)}$ and $\tilde{\mathbf{H}}_t^{(i)}$ at points where $\tilde{\mathbf{U}} = 0$: $\mathbf{z} \times \tilde{\mathbf{E}}^{(i)} = \gamma^{(i)} \tilde{\mathbf{H}}_t^{(i)}$. The exact form of $\gamma$ is not of importance at this moment: one can find an algorithm for obtaining it, as well as an explicit expression for the case of obliquely incident wave in a homogeneous medium, in Appendix \ref{app:3}. For now, all that matters is that $\gamma^{(i)}$ is completely determined by material parameters of the medium and the tangential component $\mathbf{b}$ of the wavevector. Using $\gamma$, we can write $\tilde{\mathbf{W}}^{(i)}=(\tilde{\mathbf{H}}_t^{(i)},~\mathbf{z} \times \tilde{\mathbf{E}}^{(i)})^T$ as $(P_t, ~ \gamma^{(i)})^{T} \tilde{\mathbf{H}}_t^{(i)}$. Thus, on the left-hand side of Eq. (\ref{interfaceConnection}) we have $(P_t, ~ \gamma^{(N+1)})^{T} \tilde{\mathbf{H}}_t^{(N+1)}$. The same is done to $\tilde{\mathbf{W}}^{(0)}_{z_{\text{in}}}$, with the exception that $\gamma^{(0)}$ and $\gamma^{(N+1)}$ should have different signs to ensure the same orientation of the triplet of vectors ($\mathbf{k}$, $\tilde{\mathbf{E}}$, $\tilde{\mathbf{H}}$) at $z_{\text{in}}$ and $z_{\text{f}}$. This is so, because the EM field is due to the currents lying inside the interval $(z_{\text{in}}, z_{\text{f}})$, hence, the field is emitted outwards in opposite propagation directions defined by the energy flux densities $(\tilde{\mathbf{E}}_t \times \tilde{\mathbf{H}}_t)_{z_{\text{in}}}$ and $(\tilde{\mathbf{E}}_t \times \tilde{\mathbf{H}}_t)_{z_{\text{f}}}$. If $\gamma^{(0)}$ and $\gamma^{(N+1)}$ had the same sign, the triplet ($\mathbf{k}$, $\tilde{\mathbf{E}}$, $\tilde{\mathbf{H}}$) would be right-handed at $z_{\text{f}}$ and left-handed at $z_{\text{in}}$ (or vice versa).

After multiplying Eq. (\ref{interfaceConnection}) by the block-matrix row $(\gamma^{(N+1)}, -P_t)$ from the left, we get the zero operator on the left-hand side:
\begin{widetext}
\begin{equation}
    0 = \begin{pmatrix}
				\gamma^{(N+1)}, & -P_t 
			\end{pmatrix} \Omega_{z_{\text{in}}}^{z_{\text{f}}}
		\begin{pmatrix}
			P_t \\
			-\gamma^{(0)}
		\end{pmatrix} \tilde{\mathbf{H}}_{t, z_{\text{in}}}^{(0)} + \begin{pmatrix}
				\gamma^{(N+1)}, & -P_t 
			\end{pmatrix} \int \limits_{z_{\text{in}}}^{z_{\text{f}}} \Omega_{z'}^{z_{\text{f}}} \tilde{\mathbf{U}}_{z'} \mathrm{d} z'
\end{equation}
Moving the source term to the left, it is easy to express $\tilde{\mathbf{H}}^{(0)}_{t, z_{\text{in}}}$, and after multiplying both sides by $\begin{pmatrix} P^t, & -\gamma^{(0)} \end{pmatrix}^T$ from the left, we get
\begin{equation} \label{W0Formula}
    \tilde{\mathbf{W}}^{(0)}_{z_{\text{in}}} =
		 -\begin{pmatrix}
			P_t \\
			-\gamma^{(0)}
		\end{pmatrix}
		\left[
			\begin{pmatrix}
				\gamma^{(N+1)}, & -P_t 
			\end{pmatrix} \Omega_{z_{\text{in}}}^{z_{\text{f}}}
		\begin{pmatrix}
			P_t \\
			-\gamma^{(0)}
		\end{pmatrix}
		\right]^{-}
			\begin{pmatrix}
				\gamma^{(N+1)}, & - P_t
			\end{pmatrix} \int \limits_{z_{\text{in}}}^{z_{\text{f}}} \Omega_{z'}^{z_{\text{f}}} \tilde{\mathbf{U}}_{z'} \mathrm{d} z'.
\end{equation}
\end{widetext}

Since the operator in the square brackets is degenerate in the six-dimensional space, we use the pseudoinverse operation, denoted as $[\dots]^{-}$ \cite{gantmacher1960}.  Also note that if $\tilde{\mathbf{U}}$ is identically zero, the EM field vanishes, which is expected, since we assume the field is solely due to the currents lying within $(z_{\text{in}}, z_{\text{fin}})$.

\section{\label{app:3} The $M$ operator and the surface impedance tensor}

The $M$ operator introduced in the evolution equation (\ref{basicEq}) has the block matrix form \cite{Borzdov97}
\begin{equation}
    M = 
    \begin{pmatrix}
        A & B \\
        C & D
    \end{pmatrix},
\end{equation}
where the blocks for a generic anisotropic medium are
\begin{eqnarray}
    &&A = \frac{1}{\varepsilon_{\mathbf{z}}} \mathbf{z}^{\times} \varepsilon (\mathbf{z} \otimes \mathbf{a}) - \frac{1}{\mu_{\mathbf{z}}} (\mathbf{b} \otimes \mathbf{z}) \mu P_t, \\
    &&B = - \mathbf{z}^{\times} \varepsilon \mathbf{z}^{\times} + \frac{1}{\varepsilon_{\mathbf{z}}} \mathbf{z}^{\times} \varepsilon (\mathbf{z} \otimes \mathbf{z}) \varepsilon \mathbf{z}^{\times} - \frac{1}{\mu_{\mathbf{z}}} (\mathbf{b} \otimes \mathbf{b}), \\
    &&C = - \frac{1}{\varepsilon_{\mathbf{z}}} (\mathbf{a} \otimes \mathbf{a}) + P_t \mu P_t - \frac{1}{\mu_{\mathbf{z}}} P_t \mu (\mathbf{z} \otimes \mathbf{z}) \mu P_t, \\
    &&D = - \frac{1}{\varepsilon_{\mathbf{z}}} (\mathbf{a} \otimes \mathbf{z}) \varepsilon \mathbf{z}^{\times} - \frac{1}{\mu_{\mathbf{z}}} P_t \mu (\mathbf{z} \otimes \mathbf{b}),
\end{eqnarray}
where $\varepsilon_{\mathbf{z}} \equiv \mathbf{z} \cdot (\varepsilon \mathbf{z})$, $\mu_{\mathbf{z}} \equiv \mathbf{z} \cdot (\mu \mathbf{z})$, and $\mathbf{a} \equiv \mathbf{b} \times \mathbf{z}$.

The surface impedance tensor $\gamma$ can be found as follows \cite{Borzdov97, barkovsky_furs2003}. 
Assume that we want to find $\gamma$ at some point $z$, where $\tilde{\mathbf{U}} = 0$ (remember that we only use $\gamma$ at $z_{\text{in}}$ and $z_{\text{fin}}$, where the currents vanish by assumption). Then, employing the definition $\mathbf{z} \times \tilde{\mathbf{E}}(z) = \gamma \tilde{\mathbf{H}}_t(z)$ and Eq. (\ref{basicEq}), one obtains a quadratic operator equation with respect to $\gamma$:
\begin{equation} \label{GammaEq}
    \gamma B \gamma + \gamma A - D \gamma - C = 0.
\end{equation}
Unfortunately, there is no way to obtain a concise closed-form solution for a generic anisotropic medium. A somewhat lengthy expression can be found in \cite{Borzdov97}. However, if the medium is isotropic, i.e. $\varepsilon$ and $\mu$ are scalars, then $A = D = 0$, and from Eq. (\ref{GammaEq}) it follows that $\gamma = B^{-} \sqrt{BC}$ or, explicitly, 
\begin{eqnarray} \label{GammaIsotropic}
    \gamma = \dfrac{\mu P_t - \frac{1}{\varepsilon} \mathbf{a} \otimes \mathbf{a}}{\sqrt{\varepsilon \mu - \mathbf{b}^2}}.
\end{eqnarray}
In practice, it is usually the case that the layered medium is bounded from both sides by air, vacuum, or some other isotropic medium. Thus, there is usually no need to solve the general equation (\ref{GammaEq}) for anisotropic environments, and one can use (\ref{GammaIsotropic}) instead.

\begin{figure}[h]
\includegraphics[scale=0.65]{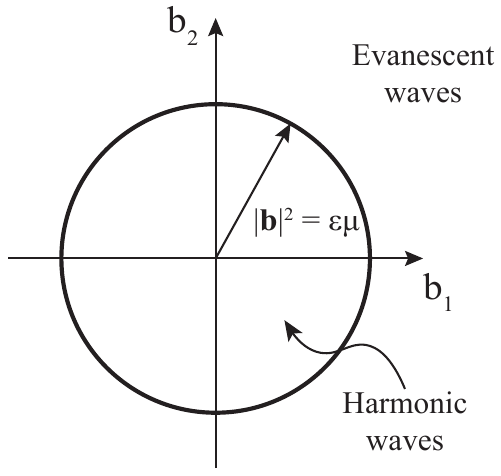}
\caption{\label{Fig6} The surface impedance tensor (\ref{GammaIsotropic}) is not defined at the circle $|\mathbf{b}|^2 = \varepsilon \mu$. The interior of the circle corresponds to harmonic waves, while the exterior corresponds to evanescent waves.}
\end{figure}

The $\gamma$ tensor in (\ref{GammaIsotropic}) is singular when $\varepsilon \mu = \mathbf{b}^2$ (Fig. \ref{Fig6}).
However, this does not lead to infinite results when integrating over all values of $\mathbf{b}$ in Eq. (\ref{DGFfin}), because $\gamma$ tensors enter the DGF in such a way that its Fourier-transformed matrix elements approach large, but finite values near the circle $\mathbf{b}^2 = \varepsilon \mu$ (see Fig. \ref{Fig2}). Still, one has to exclude the points on this circle from the integration grid, since $\gamma$ is undefined at these points. When $|\mathbf{b}|^2$ exceeds $\varepsilon \mu$, the $\gamma$ tensor becomes imaginary, which means that the Fourier amplitudes decay exponentially in the tangential plane, i.e. they become evanescent waves.

If one considers the normal incidence onto an isotropic medium, i.e. when $\mathbf{b} = 0$ and $\mathbf{H}_t = \mathbf{H}$, then the electric and magnetic fields are related by the well-known expression $\mathbf{z} \times \mathbf{E} = Z_0 \mathbf{H}_t = Z_0 \mathbf{H}$, where $Z_0 \equiv \sqrt{\mu / \varepsilon}$ is the impedance of free space. Thus, by using $\mathbf{z} \times
\tilde{\mathbf{E}}$ and $\tilde{\mathbf{H}}_t$ instead of $\tilde{\mathbf{E}}_t$ and $\tilde{\mathbf{H}}_t$ we can take $\gamma = Z_0$ when dealing with normal incidence. That is the reason why we used $\mathbf{z} \times
\tilde{\mathbf{E}}$ instead of $\tilde{\mathbf{E}}$ in our derivations.

\section{\label{app:4}Cylindrical and spherical layers}

The treatment of non-planar, e.g., cylindrical or spherical layers is ideologically the same as that of planar ones. Suppose that the properties of the medium vary only along the axis parallel to some unit vector $\mathbf{q}$, and let $\xi \equiv \mathbf{r} \cdot \mathbf{q}$ be a parameter along that axis. The general algorithm then remains the same: 1) transform the fields and currents using orthonormal basis functions suitable for the symmetry of the problem; 2) substitute the transformed fields into the Maxwell's equations in order to obtain a first-order evolution equation for tangential components of the fields; 3) solve the resulting equation for the case of a homogeneous medium and then apply the interface conditions similar to (\ref{interfaceConditions}) in order to obtain the tangential components of fields at some specific point $\xi_{\text{in}}$; 4) restore the EM field vectors in the physical space using an analog of (\ref{realWorld}) and compare the result with the definition of the DGF (\ref{DGFdef}).

A suitable orthogonal transformation for cylindrically symmetric layers would be
\begin{equation}
    \mathbf{E}(\mathbf{r}) = \sum\limits_{m = -\infty}^{\infty} \int \tilde{\mathbf{E}}_{m}(\xi, b_z) e^{i (k_0 b_z z + m \varphi)} \mathrm{d} (k_0 b_z),
\end{equation}
and for spherically symmetric layers it would be
\begin{equation}
    \mathbf{E}(\mathbf{r}) = \displaystyle \sum\limits_{l = 0}^{\infty} \sum\limits_{m = -l}^{l} F_{l m}(\theta, \varphi) \tilde{\mathbf{E}}_{l m}(\xi),
\end{equation}
where $F_{lm}$ may be expressed via spherical harmonics $Y_{lm}$ \cite{novitsky2008} as functions of polar $\theta$ and azimuthal $\varphi$ angles (same for $\tilde{\mathbf{H}}$ and $\tilde{\mathbf{j}}$). After substituting these into Maxwell's equations, one obtains the evolution equations of the same form as (\ref{basicEq}) ($\xi$ now playing the role of the $z$ coordinate used for planar media), with
\begin{widetext}
    \begin{eqnarray}
    \tilde{\mathbf{U}}_{\text{cyl}} = \frac{4\pi}{c} \left[
    \frac{1}{\varepsilon_{\mathbf{q}}}
	\begin{pmatrix}
		\mathbf{q}^{\times} \varepsilon (\mathbf{q} \otimes \mathbf{q}) \\
		-P_t \left( \mathbf{b}_{q}^{\times} + \dfrac{m}{k_0 \xi} \mathbf{e}_{\varphi}^{\times} \right) (\mathbf{q} \otimes \mathbf{q})
	\end{pmatrix} -
	\begin{pmatrix}
		\mathbf{q}^{\times} \\
		0
	\end{pmatrix} \right] ~ \tilde{\mathbf{j}}, \\
    \tilde{\mathbf{U}}_{\text{sph}} = \frac{4\pi}{c} \left[
    \frac{1}{\varepsilon_{\mathbf{q}}}
	\begin{pmatrix}
		\mathbf{q}^{\times} \varepsilon (\mathbf{q} \otimes \mathbf{q}) \\
		  P_t \dfrac{\sqrt{l(l+1)}}{k_0 \xi} \mathbf{e}_{\varphi}^{\times} (\mathbf{q} \otimes \mathbf{q})
	\end{pmatrix} -
	\begin{pmatrix}
		\mathbf{q}^{\times} \\
		0
	\end{pmatrix} \right] ~ \tilde{\mathbf{j}},
\end{eqnarray}
\end{widetext}
where $\mathbf{e}_{\varphi}$ is the unit vector along the direction of the azimuthal angle; the corresponding $M$ operators in curvilinear coordinates, along with analogs of the $V$ operator from Eq. (\ref{realWorld}), can be found in \cite{novitsky2005, novitsky2008}. One peculiarity of non-planar layers is that $M$ depends on $\xi$ \textit{even for homogeneous media}. This is basically due to the fact that such layers are compact in at least one of the tangential dimensions. So, even though the solutions to the evolution equations can still be written in the form of (\ref{homogeneousSolution}), the evolution operator $\Omega$ would not be simply an exponential of $M$.

\nocite{*}

\bibliography{apssamp}

\end{document}